\documentclass[sigconf]{acmart}

\usepackage{fancyhdr,graphicx,amsmath,amssymb}
\usepackage{graphicx,tabularx}
\usepackage{cuted}
\usepackage{multirow}
\usepackage{balance}
\usepackage{longtable}
\usepackage{subcaption}

\def\BibTeX{{\rm B\kern-.05em{\sc i\kern-.025em b}\kern-.08emT\kern-.1667em\lower.7ex\hbox{E}\kern-.125emX}}

\acmDOI{}
\acmISBN{}
\acmArticle{}
\acmPrice{}

\acmConference[recsysXfashion'19]
{Workshop on Recommender Systems in Fashion, 13th ACM Conference on Recommender Systems}
{September 20, 2019}
{Copenhagen, Denmark}
\acmYear{2019}
\copyrightyear{2019}

\renewcommand\footnotetextcopyrightpermission[1]{} 

\begin{document}

%
\title[Automated Ad Creative Generation and Ranking]{Enabling Hyper-Personalisation: Automated Ad Creative Generation and Ranking for Fashion e-Commerce}
%

\author{Sreekanth Vempati}
\affiliation{%
  \institution{Myntra Designs}
  \city{Bengaluru}
  \country{India}}
\email{sreekanth.vempati@myntra.com}

\author{Korah T Malayil}
\affiliation{%
  \institution{Myntra Designs}
  \city{Bengaluru}
  \country{India}}
\email{korah.malayil@myntra.com}

\author{Sruthi V} 
\authornote{Work done while at Myntra}
\affiliation{%
  \institution{Microsoft}
  \city{Bengaluru}
  \country{India}}
\email{vsruthi98@gmail.com}

\author{Sandeep R}
\affiliation{%
  \institution{Myntra Designs}
  \city{Bengaluru}
  \country{India}}
\email{sandeep.r@myntra.com}

%
\renewcommand{\shortauthors}{Sreekanth and Korah, et al.}

%

\begin{abstract}

Homepage is the first touch point in the customer's journey and is one of the prominent channels of revenue for many e-commerce companies. A user's attention is mostly captured by homepage banner images (also called Ads/Creatives). The set of banners shown and their design, influence the customer's interest and plays a key role in optimizing the click through rates of the banners. Presently, massive and repetitive effort is put in, to manually create aesthetically pleasing banner images. Due to the large amount of time and effort involved in this process, only a small set of banners are made live at any point. This reduces the number of banners created as well as the degree of personalization that can be achieved. This paper thus presents a method to generate creatives automatically on a large scale in a short duration. The availability of diverse banners generated helps in improving personalization as they can cater to the taste of larger audience. The focus of our paper is on generating wide variety of homepage banners that can be made as an input for user level personalization engine. Following are the main contributions of this paper: 1) We introduce and explain the need for large scale banner generation for e-commerce 2) We present on how we utilize existing deep learning based detectors which can automatically annotate the required objects/tags from the image. 3) We also propose a Genetic Algorithm based method to generate an optimal banner layout for the given image content, input components and other design constraints. 4) Further, to aid the process of picking the right set of banners, we designed a ranking method and evaluated multiple models. All our experiments have been performed on data from Myntra (\url{http://www.myntra.com}), one of the top fashion e-commerce players in India.
 




\end{abstract}

\maketitle

\begin{figure}[t]
  \includegraphics[height=60mm,keepaspectratio]{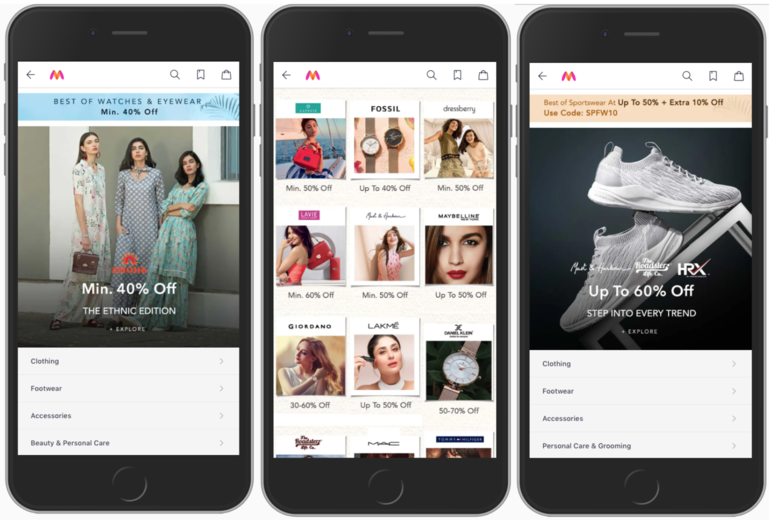}
  \includegraphics[height=57mm,keepaspectratio]{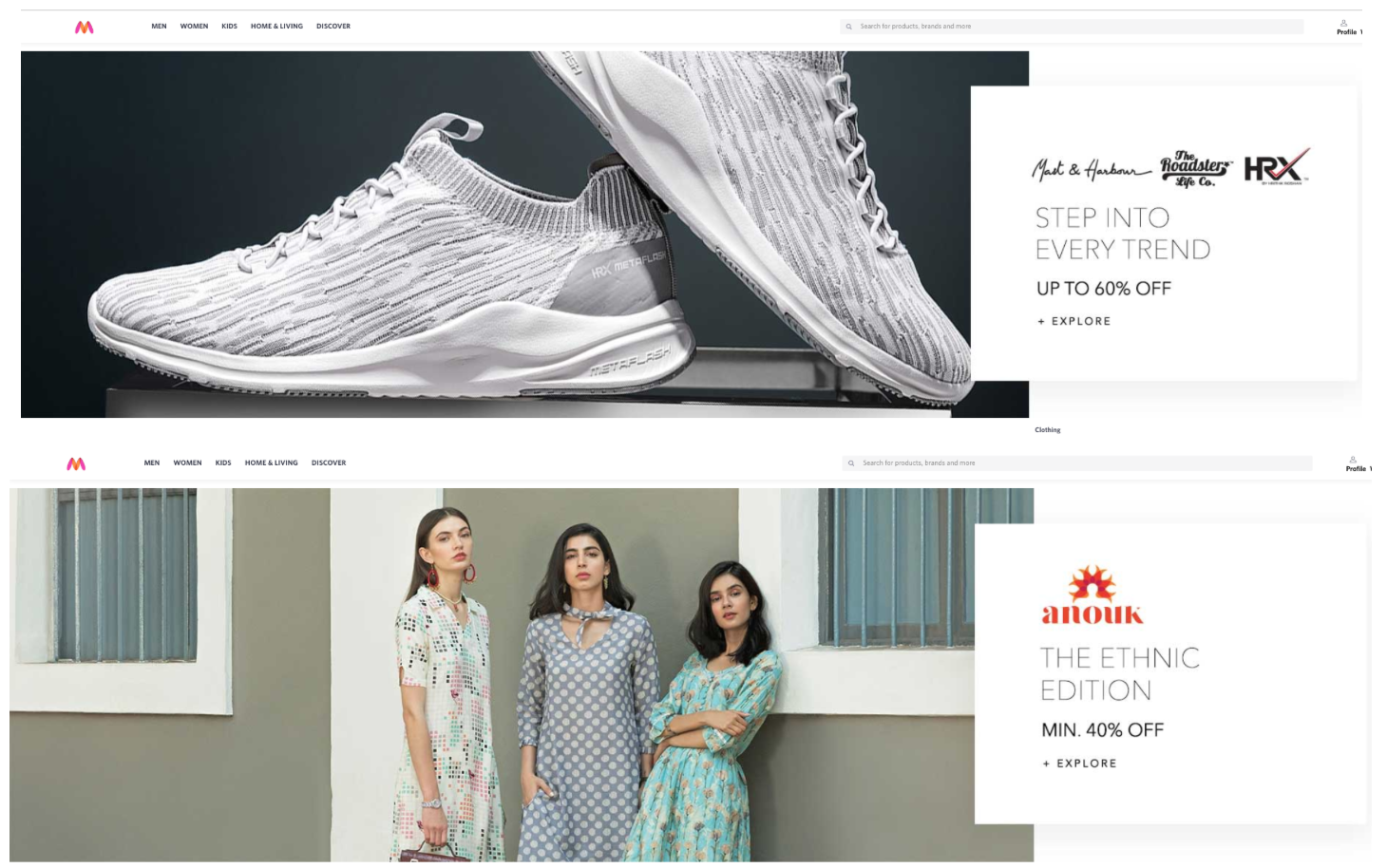}
  \caption{First row shows examples of Homepage banner/ad images on the Myntra app. Second row shows how same banner image can appear in different formats on desktop.}
  \label{fig:homepage}
\end{figure}

\section{Introduction}
In the current e-commerce era, content on the homepage plays an important role in a customer's journey for most of the e-commerce companies. 
Significant amount of the homepage on online shopping apps/websites is dedicated to banners. Example homepage banners are shown in Figure  \ref{fig:homepage}. These banners play a key role in visual communication of various messages to customers such as sale events, brand promotion, product categories and new product launches. The primary components of a banner are the image, also called a creative, and the landing page associated with it.

Any banner image is an amalgamation of an underlying theme, a background image, the brand logo in focus and affiliated text phrases. 
The text is further constituent of font colours, size and typography.
Banners essentially come down to being one particular permutation of these components which is visually appealing and can deliver the message. 
Given a requirement, a designer typically takes these various elements from a library, places these elements as per aesthetics and applies few image transformations to create a banner image.

In this paper, we show how we can automate the process that designers follow, using a set of deep learning and image processing techniques. Usually designers pick images from a large image collection called photo-shoot images. Most of these images are provided by brands or other associated entities. These images are also used on other platforms like  public hoardings, offline stores etc. Designers also use product catalogue images on few occasions. These photo-shoot images are taken in good lighting conditions showcasing highlights of the brands.

Due to the notably large amount of manual hours invested in creating these banners, only a small set is made live at any point of time. 
This reduces both the variety of output produced and the degree of personalization that can be achieved. 
But if the range of available banners increases, satisfying the personal taste of a larger audience can be achieved. 
From a wider spectrum of options, we can now accomplish targeted banners that cater to different sectors of customers, instead of generic ones made for all. In this paper, we do not focus on the user personalization engine, which is already internally built, but we focus on large scale generation of the inputs to this engine.

We present a method which generates banner images using a library of design elements. Examples of design elements include background content image, text phrases, logo etc., One of the key design elements is the underlying layout of the banner which determines the spatial arrangement of various design elements like text, logo etc., A layout can be used to describe a group of banners, and can also be used to create new banners with the same blueprint. We use a Genetic algorithm based method which can generate a banner layout given a background image, logo and text. It takes into account, design considerations like the symmetry, overlap between elements, distance between elements etc for assessing layout quality.  Overall, we can generate the banners using a library of photoshoot images and tag lines for any given themes. As input, we make use of photo-shoot images that are available to the designers in e-commerce companies. We further present an offline process to rank these generated banners. This method constitutes of machine learning models built on banner meta-data.

The use cases of this proposed method are not restricted to e-commerce websites only. It can further be extended to automating banners for social messaging platforms and movies or online video content providers. 

In the next sections, we discuss some of the related work to this problem and then we talk about the method for creating the banners, touch upon few challenges and solutions. Further, we present and evaluate the methods for evaluating the banners generated using different design elements.

\vspace{-3mm}
\section{Related Work}
As this is an attempt to solve a new problem, there is very limited related work available for the same. We do have some research being done to solve similar problems in other domains. In the past, there have been papers which solve the problem of automated layout design for photo-book\cite{photobook} and magazine cover \cite{magazinedesign}.

Photo-book design system \cite{photobook} was built to automatically generate photo compositions such as collages or photo-books using design and layout principles. In their paper, images and text elements are pre-processed and then content is distributed across pages using pre-defined rules. Then, a layout is generated using a set of design principles. Further, genetic algorithm is used to evaluate the layout, whose fitness function takes into account the borders, text elements and overall visual balance of the page.


There is some work on generating new layouts and also transferring the layout using an example \cite{layoutgan}\cite{adobepaper}\cite{layout_alibaba_paper}. In \cite{layoutgan}, Generative Adversarial Networks are used to generate new layouts for UX design and clip-art generation. The generator takes randomly placed 2D elements and produce a realistic layout. These elements are represented as class probabilities and bounding boxes. A CNN based discriminator is used as the aim was to be similar to the human eye and spatial patterns could be extracted. 

Alibaba's Luban \cite{luban}\cite{lubanBlog} is the most relevant one to our work, but we do not have any published work regarding this. As per the blog, the design team established a library of design elements and allowed Luban to extract and cluster features from raw images. Then, given a blank canvas, it places elements on it in a random fashion and then uses reinforcement learning to learn the good designs.

\cite{layout_alibaba_paper} tackles the problem of multi-size and multi-style designs, i.e. modifying a single design to suit multiple styles and sizes. Automation of layout design by optimizing an energy function based the fitness of a layout style which measures factors such as margin, relative position, etc. In \cite{adobepaper}, a new banner is being generated based on a given examples using energy function optimization.


Another energy based model is built by \cite{single_page_layout} targeting single page graphic designs. The images are analyzed to get hidden variables such as importance of different elements and alignment. The energy based model is then  based on positions of different elements, balance, white space, overlap area, etc. Design re-targeting is also presented, i.e, transferring the same design to a different aspect ratio. We have adopted some of the energy terms in generating the layout in our work.

In the movies domain, Netflix\cite{netflix_paper} blogs talk about generating art work which is personalized to the user. The focus here is primarily on finding the best thumbnail for a given movie and user by using reinforcement learning approaches. 

For predicting the Click-Through-Rate (CTR) of online advertisements, \cite{explore_ads} has trained various models on a large dataset of images. In Deep CTR system \cite{deepctr}, a deep neural network is proposed in which uses convolution layers to automatically extract representative
visual features from images, and nonlinear CTR features
are then learned from visual features and other contextual
features by using fully-connected layers. There is also work on finding the quality of native ads using image features \cite{nativequality}.

As our aim is to automatically generate a large number of visually appealing banners from a library of photoshoot images, brand logos and associated text, we have adopted a multi-pronged approach in solving various aspects of the problem. The original images are annotated and re-sized. The best possible layout is generated and post processing steps are performed. The final generated creatives are ranked according to their aesthetic appeal and historical performance. The entire approach is explained further.

\begin{figure}[ht]
\includegraphics[height=100mm,keepaspectratio]{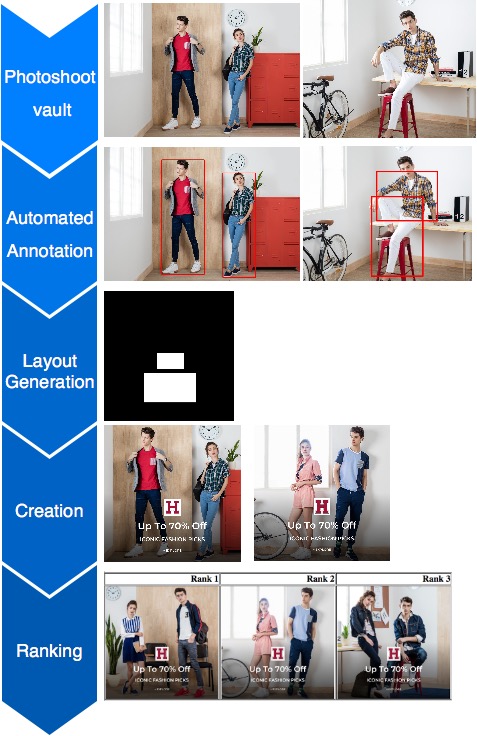}
  \captionof{figure}{End-to-end pipeline for automated creation of banners}
  \label{fig:end_to_end_pipe}
\end{figure}


\section{Methodology}

Given that the integral constituents of a banner are the image, logo, text and additional elements, structuring them optimally results in the end product. This structure is defined by a layout, which is essentially the positions and sizes of the various components. Hence the input to automated generation would primarily be a layout and also the image in focus, the brand's logo, text and other associated details. We could use both human provided layout as well as generate a layout based on the given photoshoot image and design elements. We present a layout generation method in the further sections. Different layouts are generated and the best ones are utilized for final creatives generation.\par

For automation to be possible, we need automated tags to pick the right image for the right theme/purpose. A bare photo-shoot image is composed of various fashion articles spread across it. Examples of fashion articles are ``Men-Tshirts", ``Women-Jeans" etc.,  Information regarding this is required for filtering and selecting appropriate ones. Thus the first step would be to perform large scale automated annotation of all images and tag each of them with relevant data. \par

Once annotation is complete, this newly obtained information is given as input to the layout generation module and further to creation module. The region of interest is extracted, and the different elements are stitched together according to the layout specifications. \par
The final result produced is a large number of banners for the given theme. Since only a few from this pool of options would actually be required , these are further re-ranked by a model built on historical data.

End to end steps involved in the banner creation and ranking can be found in the Figure \ref{fig:end_to_end_pipe}.

\subsection{\textbf{Automated annotation of Photo-shoot images}}

Automated annotation involves extracting the meta-data of an image. Simpler attributes like brand name and season name are given as a label. For all the other attributes, we need to visual understanding of the image which is done by using a set of detectors for each attribute. The constituents of the images are tagged based on categories or bounding boxes. A bounding box is a rectangle that completely encloses the object in focus and is described by coordinate points. 

The different aspects of annotation ranges from the significant objects such as people or the main article, along with their types to the secondary level details such as a person's gender, the type of scenery of the image, and the number of articles present in each category. We explain more details for each of the aspects below.

\subsubsection{\textbf{Object and Person Detection}}
The various objects present in the image was detected using the MaskRCNN\cite{maskrcnn} object detector.\par 
MaskRCNN was built for the purpose of instance segmentation, which is essentially the task of detecting and delineating distinct objects of interest in an image. The state of the art instance segmentation techniques are R-CNN\cite{rcnn}, Fast-RCNN\cite{fasterrcnn}, Faster-RCNN\cite{fasterrcnn} and MaskRCNN\cite{maskrcnn}. MaskRCNN was chosen as it addresses the issues of Faster-RCNN and also provides pixel-to-pixel alignment.

In particular, detecting people is of more interest to us as most of the times the article in focus is on or near them. Thus images were tagged with the bounding boxes of people present and additional information such as total number of people, dominant person, etc. We have used pre-trained detector for our person/object annotations.

\subsubsection{\textbf{Fashion Category Detection}}
\begin{figure}[t]
  \includegraphics[height=47mm,keepaspectratio]{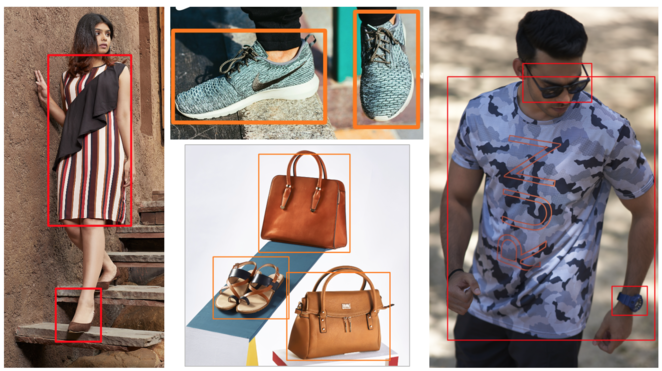}
  \captionof{figure}{Example images showing how the Fashion Categories detected from the photoshoot images. We can see the detected objects includes regions covering wide range of categories such as footwear, clothing, bags, etc.}
  \label{fig:detections}
\end{figure}

Though detecting the people in image narrows down the region of interest, the actual focus is always on the product that the banner was originally meant to be created for. Tagging this product will help us give significant importance to it. Thus, to identify the various fashion categories present in an image, an in-house detector was used. This fashion category detector was built using Mask RCNN architecture\cite{maskrcnn} and was trained on fashion specific categories like shoes, watches, etc., The training data for this detector contained manually labelled bounding boxes and class labels for the fashion categories. Example detections are illustrated in Figure \ref{fig:detections}. The mean average precision of this detector (mAP) is 67.9\%. This value is inline with the mAP obtained for the famous Pascal VOC 2007 dataset which is 63.4\%. 

This detector provides a bounding box of the category along with it's type. The types of categories include top-wear, shoes, watches, etc. Additionally, the image was also tagged with the dominant article present in terms of area. This would later be useful in filtering images based on the articles present. 


\subsubsection{\textbf{Gender and Face Detection}}
Apart from the entire person's body, detecting the face will be of more use. This is due to the fact that in certain cases it is okay to overlap other elements on a person's body, but the design elements should not be present on the face. Tagging the gender of the people present will again help filter images based on any particular theme requirement. A CNN-based gender detection model \cite{gender_detection} trained on the IMDB-Wiki dataset was used for this purpose.

\subsubsection{\textbf{Scene Detection}}
Using the scene detector\cite{scene_paper}, we obtain the various attributes of the background of an image. The categories of scenes include indoor, outdoor as well as details like restaurant, garden, etc and features such as lighting level and man-made area. This level of annotation will help filtering images for both theme based creatives generation and for better personalization.

\subsubsection{\textbf{Text Detection:}}
We perform text detection on photoshoot images so as to remove few images which have too much text area in the image and are not suitable for generation of creatives. Text was detected using the OpenCV\cite{opencv} East Text detector.

\subsection{Layout Generation}
A layout is defined as the set of positions/placements for each of the design elements like ``brand logo", ``text callouts" etc., on a given content image consisting of people and/or objects along with their bounding boxes. A layout $L$ can be defined as $\{{\theta}_1,{\theta}_1 \dots {\theta}_n \}$ where ${\theta}_i$ represents the co-ordinates of the bounding box for the $i^{th}$ design element. Our objective is to find the co-ordinates ${\theta}_{i}$ which form the layout with highest aesthetic value. 

\begin{figure*}[ht]
    \centering
    \includegraphics[width=0.55\textwidth,keepaspectratio]{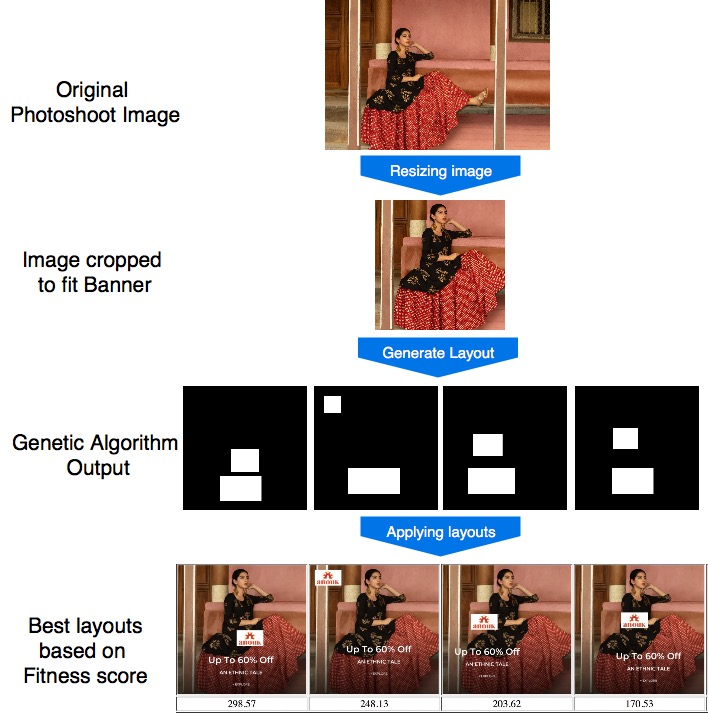}
    \captionof{figure}{The steps involved in layout generation, and applying them on creatives. The generated layouts and their respective fitness scores are observed. The scores are found to be congruent with the aesthetic value of the banner}
    \label{fig:pipeline}
\end{figure*}

Layout generation involves evaluating aesthetic compatibility for all combinations of possible locations to properly place and scale the design elements in the banner. Since there are very large number of combinations of the possible coordinates, it is time-consuming to compute to a feasible solution out of all the combinations. For this purpose, we have used Genetic algorithms, as they are suitable for problems involving large combinations of variables. Genetic algorithms have been proven to help in converging quickly to a workable solution and the random mutations also take care of generating new unseen layouts.


Genetic algorithm \cite{Genetic_algo_paper} simulates the natural process of evolution and uses the following techniques to reach the best possible/fittest solution. It starts with an initial population and performs the following steps to reach the best possible solution. In our case, each of the population corresponds to one layout configuration of all the design elements. \par

a) Selection : Chooses the best parents to produce the next generation from a population \par
b) Crossover:  Combines two parents to create new individuals. In our case, this involves swapping coordinates between the various design elements like text boxes/logo. \par

c) Mutation: Randomly changes genes/points on individuals to create new individuals. This helps in evaluating new coordinate points. \par

d) Fitness Function: Uses a fitness score to evaluate the population. Individuals having a higher fitness score are chosen as the next parents.  \par

The Distributed Evolutionary Algorithms in Python (DEAP) library \cite{DEAP_JMLR2012} was used for the implementation.

The algorithm generates a random population with $x$ and $y$ coordinates for the logo and text bounding boxes. The bounding boxes of the persons and objects in the photoshoot image is considered as fixed. These coordinates are the inputs for the model. The algorithm performs a series of selection, crossovers and mutations on the logo and text coordinates to come up with the best solution based on the fitness function and the constraints provided. The fitness function incorporates the fundamentals of graphic design by aggregating scores for each of the design aspects. Once a specific number of generations are produced, the individual corresponding to the best score is chosen as the output.  


The final fitness/energy score for a layout, $E(X, \theta)$, is the weighted sum of individual  scores, $E_i(X,\theta)$. One such layout assessment is showcased in \cite{single_page_layout}. The weights for the individual fitness scores were obtained by regression of these scores on the CTR of historical banners. We've used CTR, as it helps in decoding user preferences, there by helping us in mapping the weights for different design aspects to user preferences.

\begin{equation*}
E(X, \theta) =  \sum_{i=1}^{n}w_iE_i(X,\theta)
\end{equation*}
X is the background layout which includes the $x$ and $y$ positions of the bounding box for the person/object in the image. It is denoted as an array $[x_{left}, y_{top}, x_{right}, y_{bottom}]$ which corresponds to the top-left and bottom-right coordinates of the bounding box. ${\theta}$ represents the coordinates for each element and hence is an array of 4 elements. $w_i$ represents the weight for the $i^{th}$ fitness term. Key individual fitness scores are explained below. Note that the overall fitness function can be easily modified to incorporate more design rules.

\paragraph{Alignment:}
Alignment is one of the core designing considerations while creating a banner. We calculate the misalignment metric which penalizes layouts where the elements are misaligned. For cases with left alignment, we have a lower penalization. 

\paragraph{Overlap:}
Overlapping of any two of the elements, significantly reduces the aesthetic value of the banner. We calculate the overlap percentage for all pairs of elements and penalize them in the fitness score.
\begin{equation*}
Overlap \% = \frac{Area_{overlap}}{Area_{total}}
\end{equation*}

\paragraph{Distance between elements:}
Even in cases of zero overlap between the elements, there can be cases where they are placed very close to each other. This is especially discomforting when the logo or text are placed very close to the face of the person or the important region of an object in the background image. Hence layouts with elements placed farther apart are preferred. The euclidean distance is calculated between pairs and added:

\begin{equation*}
  \begin{gathered}
    Distance(i, j) = \sqrt{(x_i - x_j)^2 + (y_i - y_j)^2}
    \\  \forall (i,j) \; i \neq j 
    \\ i,j \in \{Person, Logo, Text, \dots \}
    \end{gathered}
\end{equation*}

\paragraph{Symmetry:}
Symmetry of the layout is a key factor in stabilizing the overall balance of the final layout. To account for this, we calculate the asymmetry for all elements in the image layout and add this as a penalization term in the fitness score.

\begin{equation*}
{X_{center} = \frac{(X_{left} + X_{right})}{2}}
\end{equation*}

\begin{equation*}
{Asymmetry_{horizontal} = |2 *  X_{center} - Width_{layout}|}
\end{equation*}


\paragraph{Constraints:}
To make sure that all elements are of reasonable size and not too large, the bounding boxes for the elements are assigned a buffer  region. All layouts where the dimensions fall outside this region, we term those as in-feasible solutions. 

\paragraph{Qualitative Evaluation:}
The generated layouts are sorted according to their fitness scores and the top one is selected. We carried out an internal survey asking users to identify the designer created layouts and the machine generated ones. 
It was observed that 60\% of the people were not able to distinguish between them. An example result is illustrated in the Figure \ref{fig:pipeline}.

For illustration purposes, We have tried to maintain uniformity within single brand/category level creatives and hence the layout that performs best on majority of the cases in further examples.

\subsection{\textbf{Creative Generation}}
Combining everything together, creative generation involves following steps for a given library of photoshoot images with annotations, brand logos, text callouts.

\begin{itemize}
\item Filtering the relevant photoshoot images for the given brand or category using automated tags. Brand name is already provided for the images as labels.
\item Automatic cropping of the selected photoshoot images to find the region of interest using annotations.
\item Generating best layouts for the given cropped image, logo and text call-outs. 
\item Overlaying of the different design elements as per the layout specification. Details on how the text is selected and overlaid along with few post-processing steps are explained below.
\end{itemize}

\begin{figure*}[ht]
    \centering
    \includegraphics[width=1.0\textwidth,keepaspectratio]{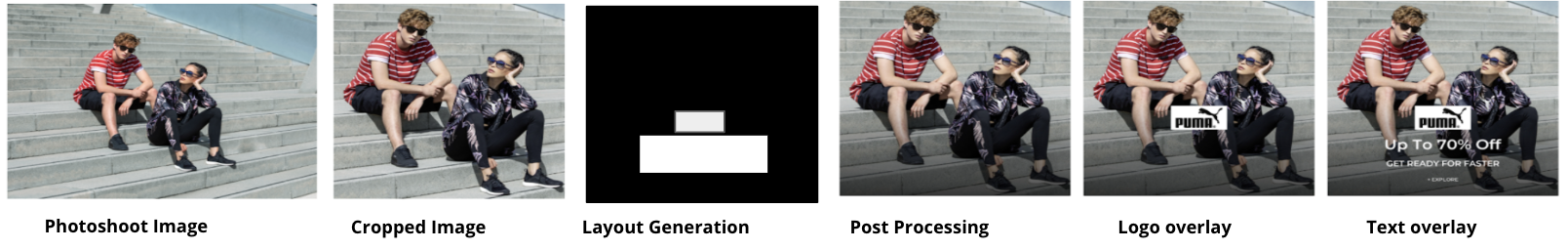}
    \captionof{figure}{Above figure illustrates the prominent steps involved in creating the banner creative. Given a photoshoot image, it is automatically cropped according to the given requirement of size and category, then best layout is computed using the genetic algorithm based method. After applying image post processing steps, design elements like logo and text call-outs are placed as per the layout specifications.}
    \label{fig:prominent_steps}
\end{figure*}

All the steps involved in creative generation are illustrated for an example in the Figure \ref{fig:prominent_steps}.

\subsubsection{\textbf{Overlaying text callouts}}
The text on a banner signifies the core message and intention for creating it. Be it a sale or a new arrival, catchy phrases are used to attract the customer. We have a collection of text phrases for various new launches, brand call-outs, sale events. We select the appropriate text to be overlayed from this collection.

\paragraph{Text Formatting}
We have a few pre-defined standard text colours, size and fonts based on banner images served on the platform in the past. These were obtained using certain existing design rules, such as golden ratio and the rule-of-thirds. 
The golden ratio(divine proportion) acts as a typography ratio that relates font size, line height, and line width in an aesthetically pleasing way. \par 

\paragraph{Post Processing}
While adding text on an image, we have to ensure that the font color is appropriate. To allow even lighter font colors to be visible, a slight darker gradient is applied on the image. 




\subsubsection{\textbf{Baseline Approach:}}
As a baseline approach for generating creatives, we can crop the center region of the given photoshoot image with required aspect ratio. Further steps include pasting brand logo and text in the same layout as used in the creative generated. Some image processing is done onto the banner creative. Note that this approach doesn't consider regions of any objects/people present in the image. Results using this approach can be seen in Table \ref{tab:baseline}.


\begin{figure}[ht]
  \includegraphics[height=40mm,keepaspectratio]{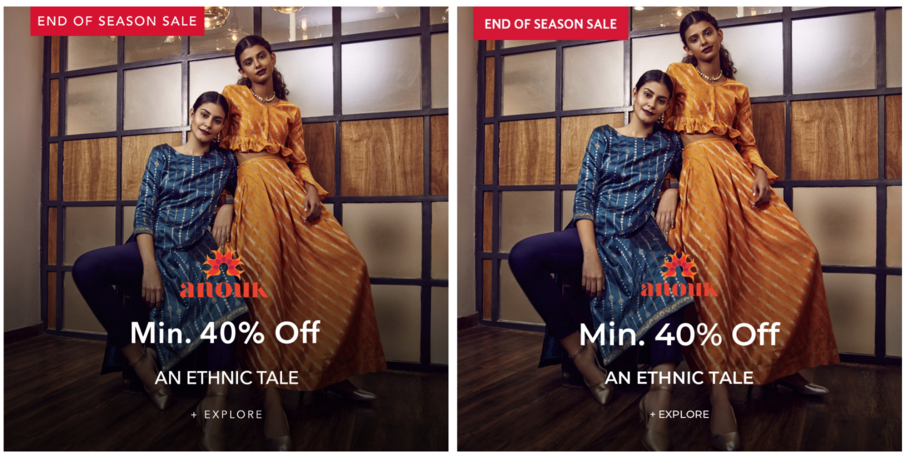}
  \includegraphics[height=40mm,keepaspectratio]{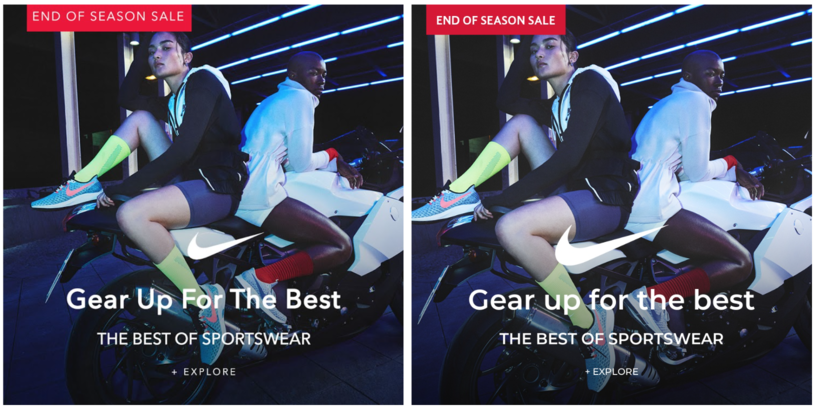}
  \includegraphics[height=40mm,keepaspectratio]{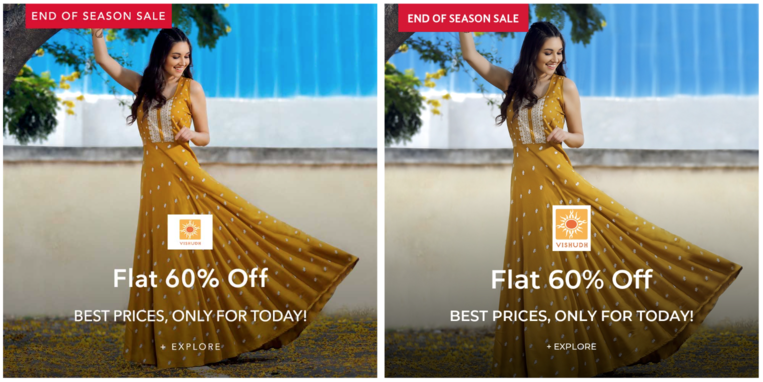}
  \captionof{figure}{The image on the left was created by a designer while the one on the right was automatically generated.}
  \label{fig:manual_new}
\end{figure}

\subsubsection{\textbf{Qualitative Evaluation of generated Creatives:}}
\label{section:generated} 
\begin{figure}[hh]
  \includegraphics[width=0.8\linewidth,keepaspectratio]{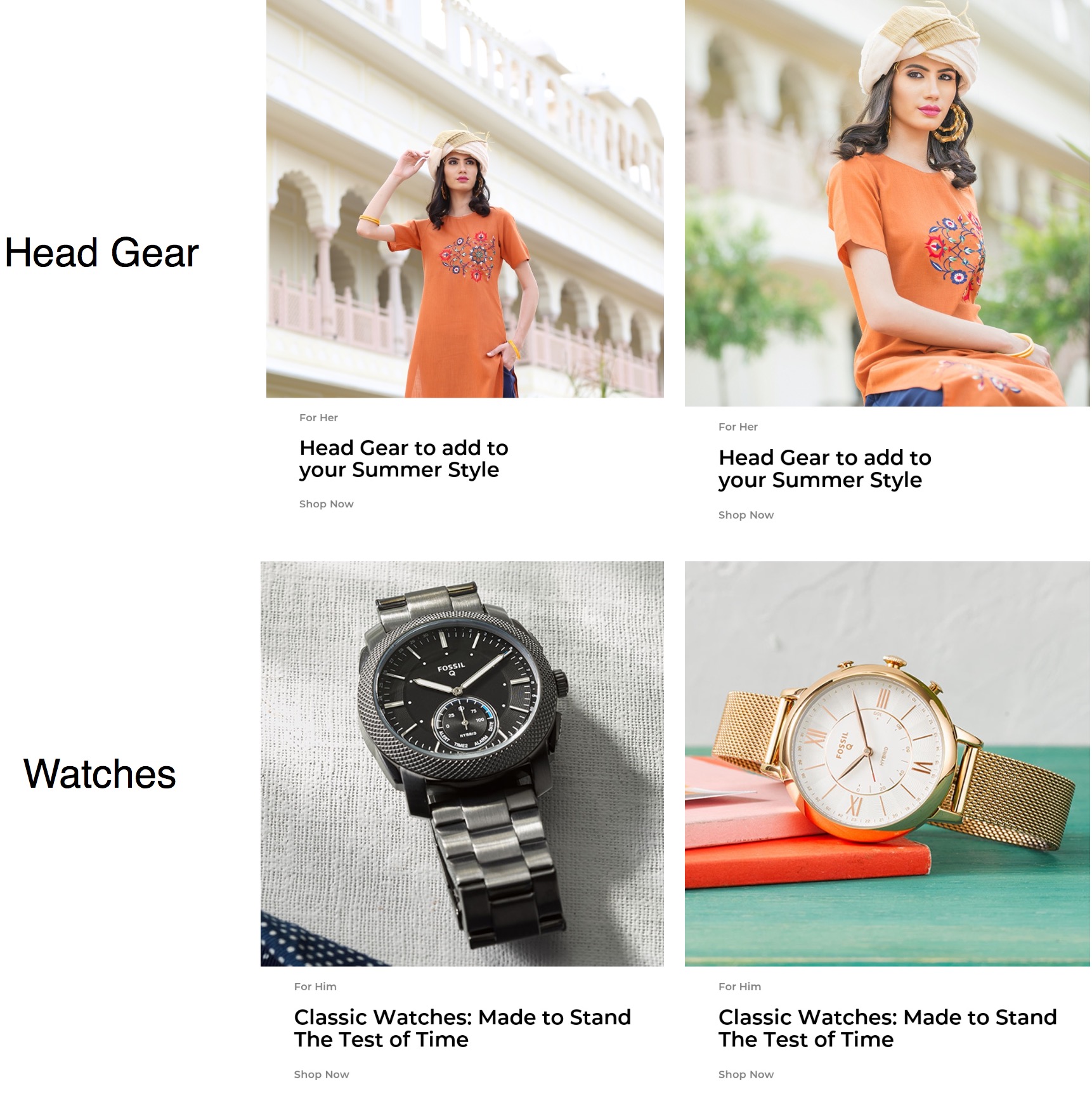}
  \captionof{figure}{Creatives generated for Article types. First row correspond to the ``head gear" category and second row shows creatives generated for ``watches" category.}
  \label{fig:article}
\end{figure}

Figure \ref{fig:manual_new} shows examples of banners which were generated by designers and our approach. Figure \ref{fig:article} demonstrates how article-type based creatives are generated. To evaluate the goodness of the generated creatives w.r.t designer created banners, we had conducted an internal survey. In the survey, a user was presented with two images, a designer created banner and automated banner. We found out that only 45\% of the people were able to judge correctly, showing that we were able to create banners which are realistic. 


\subsection{Ranking creatives}
So far, we have explained ways of automatically generating the banners. Out of these numerous banners generated, we would prefer the superior ones to go live. In order to eliminate the manual effort put in picking these banners, we have designed a ranking methodology based on historical live data-set which had images along with their associated Click-Through-Rate (CTR) as labels. This methodology can be used to rank manual as well as automatically generated creatives. We consider CTR as a signal for goodness of a creative. Note that we do not have any ground truth labels for the generated creatives. We test the goodness of our creatives with the help of the ML model trained on the historical dataset (creatives which are manually designed) which already has a label in terms of CTR. We have trained different models on this dataset to predict CTR given a generated creative.

Note that the layout generation algorithm explained in earlier sections does the job of finding best layout where as the method explained in this section helps in ranking all creatives and considers features which not only explain about layout, but also the content in the image and its historical performance.

\paragraph{\textbf{Feature Engineering:}} The model was built on both image embeddings, explicit features designed from the layout of the image and a set of features determining aesthetic score of the image.

\paragraph{VGG Embeddings}
The VGG convolutional network\cite{vgg} that has been trained on the ImageNet database provides embeddings for an image of dimension $4096$. Input to the network requires the image to be re-sized to $224 x 224$. 

\paragraph{Layout Extracted Features}
Using the various annotation means used, we can  engineer features from the different components of an image. 

\begin{enumerate}
    \item Position specific features : The coordinates of bounding boxes for people, text, faces and articles.
    \item Area : The relative area percentage covered by each of the dominant objects. 
    \item Gender : Number of women present, Number of men present, Total number of people present.
    \item Category Type : The types of articles detected are one hot encoded (present or not) . Types include topwear, bottomwear, watches, etc.
    \item Environment Type : Indoor or outdoor background setting.
    \item Scene Categories and Attributes : Frequently occurring categories (restaurant, garden, etc) and attributes(lighting,man-made, etc.) were picked and one-hot encoded. 
    \item Overlapping objects : When text is overlayed on the image, it will overlap on the existing components such as a articles or a person's body. This overlap is tolerable as long as the main article of focus or a face is not hidden. To account for this, the relative area percentage of overlap between each of the following components are calculated :
    \begin{itemize}
        \item Text regions and Faces
        \item Text regions and People
        \item Text regions and Articles
    \end{itemize}
    \item Text Quadrants : One hot encoded information for every quadrant if it contains a text component or not.
    
\end{enumerate}

\paragraph{Aesthetic Features:}
We have used Neural Image Assessment (NIMA scores)\cite{nima_google} which computes scores representing aesthetics of an image. This is obtained by training a deep CNN on a dataset containing images along with human judgment on aesthetics. In our experiments, we have obtained the score by using this publicly available pre-trained model \cite{nima_google}. For a given image, the aesthetic scores predicted by this model was used as additional feature.

\paragraph{\textbf{Ranking models:}}
Apart from the simple Logistic Regression model, the tree based classifier were chosen as they are popular methods for CTR prediction problems. Note that other methods based on deep learning \cite{deepctr} could further improve the ranking methodology. 

Here are the methods that we have experimented along with the optimal parameters picked. 
\begin{enumerate}
    \item Logistic Regression
    \item Decision Trees 
    \item Random Forest Classifier 
\end{enumerate}

As there are fewer clicks compared, we have balanced the data by providing higher weights to samples with clicks. 

\paragraph{\textbf{Ad Personalization Engine:}}
In order to provide a personalized experience, relevant ads are being chosen and shown to the user. For this purpose, all the active ads will be ranked using prediction scores of a click-prediction model. This model is trained on historical click-stream data with user and ad features as input variables. This model is already deployed on production and is not the focus of this work.


\section{Experiments and Results}
In order to evaluate the goodness of the proposed ranking approach, we have conducted few offline experiments. To evaluate the overall approach, we have performed an online A/B test. Details are explained below. 

\paragraph{Dataset} All the images and data used for our experiments are taken from Myntra (\url{http://www.myntra.com}), one of the top fashion e-commerce players. We have trained trained the models on nearly 20,000 existing production banner images which already have labels. We perform the feature extraction on all the images in the dataset and then use the target variable as is\_clicked. The models were trained on 75\% of the data and tested on 25\%. 

\begin{figure}[hb]
  \includegraphics[height=50mm,keepaspectratio]{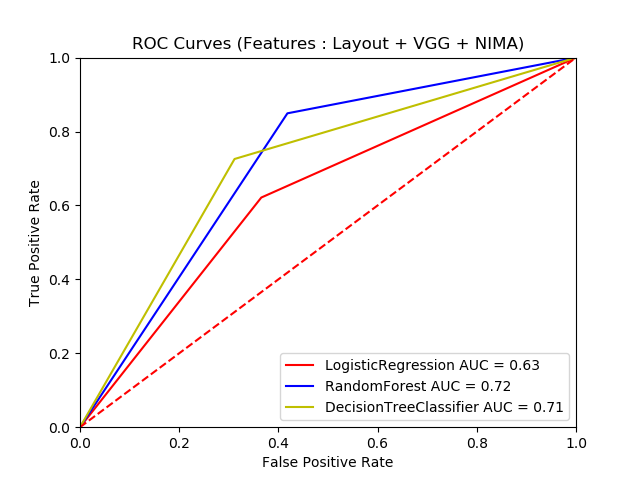}
  \captionof{figure}{ROC curves for the different classifiers on the combined feature set : VGG embeddings, NIMA and Layout extracted features.}
  \label{fig:manual}
\end{figure}


\begin{figure*}
    \centering
    \includegraphics[width=0.7\textwidth,keepaspectratio]{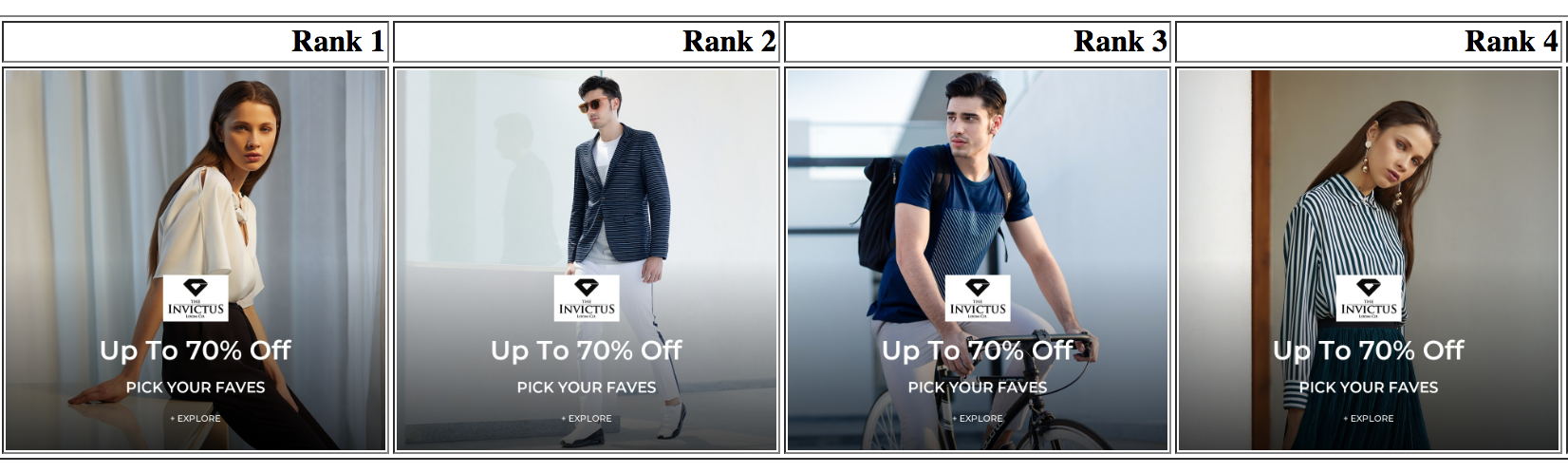}
    \captionof{figure}{Examples of images with high predicted CTR for the ones generated under a single brand.}
    \label{fig:good_ctr}
\end{figure*}

\begin{figure*}
    \centering
    \includegraphics[width=0.7\textwidth,keepaspectratio]{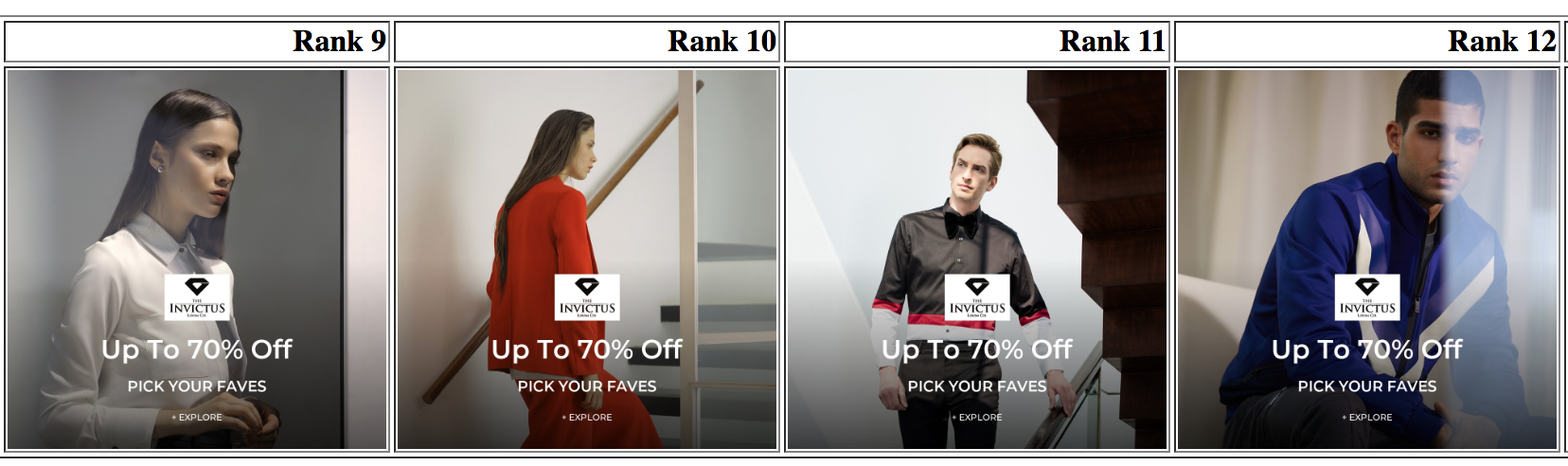}
    \captionof{figure}{Examples of images with low predicted CTR for the ones generated under the same brand. We found that ranking model is inline with human judgment.}
    \label{fig:bad_ctr}
\end{figure*}

\paragraph{Evaluation Metrics} Classification models were evaluated using both AUC (Area Under Curve) and NDCG (Normalized Discounted Cumulative Gain) \cite{ndcg}.
NDCG is a standard information retrieval measure\cite{ndcg} used for evaluating the goodness of ranking a set. For computing NDCG, we have used CTR as relevance score. Evaluation Metrics on test data using different models is presented in the Tables \ref{table:features} and \ref{table:models}.

\begin{table}[]
\begin{tabular}{|l|l|l|}
\hline
\textbf{Features}                           & \textbf{AUC} & \textbf{NDCG} \\ \hline
VGG Embeddings                              & 0.71         & 0.17          \\ \hline
Layout extracted features                   & 0.74         & 0.14          \\ \hline
NIMA                                        & 0.71         & 0.24          \\ \hline
VGG Embeddings + Layout Features            & 0.71         & 0.17          \\ \hline
Layout extracted features + NIMA            & 0.72         & 0.56          \\ \hline
VGG + Layout features + NIMA                & 0.71         & 0.22          \\ \hline
\end{tabular}
\caption{Evaluation metrics for different feature sets experimented on best performing model (Random Forests)}
\label{table:features}
\end{table}


\begin{table}[]
\begin{tabular}{|l|l|l|}
\hline
\textbf{Model}      & \textbf{AUC} & \textbf{NDCG} \\ \hline
Logistic Regression & 0.60         & 0.03          \\ \hline
Decision Trees      & 0.70         & 0.05          \\ \hline
Random Forests      & 0.71         & 0.22          \\ \hline
\end{tabular}
\caption{Evaluation metrics for different models experimented on combined feature set i.e., VGG embeddings, NIMA and Layout extracted features}
\label{table:models}
\end{table}

\paragraph{Quantitative Evaluation} 
From the various models and features experimented with, the promising results are shown in Tables \ref{table:features} and \ref{table:models}. Apart from training on VGG embeddings, NIMA scores\cite{nima_google} and layout extracted features individually, a combination of all was also attempted. Since the best performing model was the Random Forests Classifier, the results for it are present in Table \ref{table:features}. Performance metrics using different models for combined feature set is present in Table \ref{table:models}. We can see that Random Forests using Layout and NIMA features gives the best NDCG. ROC curves for different classifiers using all the feature sets can be see in Figure \ref{fig:manual}.


It is also interesting to observe that when the model was trained only on the layout extracted features, the most important features were : area of the image, overlap area between text and objects, certain text region coordinates, overlap area between text and people, etc. This further reiterates the fact that the position and orientation of text defines the layout of an image, and it is useful to generate banners without assumptions about their positions. 

\paragraph{Online evaluation:}
We have also performed an online A/B test of the auto-generated creatives along with the manually designed creatives for the same brands. As the main hypothesis was that having multiple options increases personalization, we had a larger set of automated banners in the test set compared to the control set. For both buckets of the A/B test (equal proportion of user traffic), we have used the same personalisation engine that considers both user preferences and banner features. The control group contains manually designed banners and the test group contains automated banners. The results have shown that CTR has increased by 72\% for the test set compared to control set (relative increment), with high statistical significance. 

\paragraph{Qualitative Evaluation} 

When the model trained on historical data was used to predict CTR on the newly generated creatives, the results were quite similar to a what the human eye would observe. \par

Figure \ref{fig:good_ctr} is an example of images that have high predicted CTR. This seems to be a meaningful outcome as these images have good color contrast ratio, optimal placement of the components and visible text. In Figure \ref{fig:bad_ctr}, we notice how the images with poor lighting, faces turned away and ones with unnecessary extra padding space are all pushed down the ranking scale due to much lower predicted CTR.

\section{Conclusion}

In this paper, we have presented a method to generate banner images in an automatic manner on a large scale. This helps in reducing the massive effort and manual hours spent by designers currently and also to produce a wider variety of options to pick from. The broader spectrum of banners will help in catering to wide spectrum of users, instead of showing common banners to all users. We have presented end to end steps involved in generating creatives given few input constraints using automated annotation of photoshoot images. We described a genetic algorithm based method to generate a layout given cropped image and other design elements. 

We have shown how a ranking model can be trained on historical banners to rank these generated creatives by predicting their CTR. We observed that when the best performing model was tested on these automatically produced banners, the ranking results was very similar to what a human would've picked, with well positioned, optimal images having higher CTR than the rest. Apart from this offline method of ranking them, future work would be to perform online ranking via reinforcement learning, which will also further boost the personalisation by showing the right set of banners from the vast amount of banners created.



%

%
\bibliographystyle{ACM-Reference-Format}
\bibliography{sample-sigconf}


\begin{thebibliography}{23}


\ifx \showCODEN    \undefined \def \showCODEN     #1{\unskip}     \fi
\ifx \showDOI      \undefined \def \showDOI       #1{#1}\fi
\ifx \showISBNx    \undefined \def \showISBNx     #1{\unskip}     \fi
\ifx \showISBNxiii \undefined \def \showISBNxiii  #1{\unskip}     \fi
\ifx \showISSN     \undefined \def \showISSN      #1{\unskip}     \fi
\ifx \showLCCN     \undefined \def \showLCCN      #1{\unskip}     \fi
\ifx \shownote     \undefined \def \shownote      #1{#1}          \fi
\ifx \showarticletitle \undefined \def \showarticletitle #1{#1}   \fi
\ifx \showURL      \undefined \def \showURL       {\relax}        \fi
\providecommand\bibfield[2]{#2}
\providecommand\bibinfo[2]{#2}
\providecommand\natexlab[1]{#1}
\providecommand\showeprint[2][]{arXiv:#2}

\bibitem[\protect\citeauthoryear{Amat, Chandrashekar, Jebara, and
  Basilico}{Amat et~al\mbox{.}}{2018}]%
        {netflix_paper}
\bibfield{author}{\bibinfo{person}{Fernando Amat}, \bibinfo{person}{Ashok
  Chandrashekar}, \bibinfo{person}{Tony Jebara}, {and} \bibinfo{person}{Justin
  Basilico}.} \bibinfo{year}{2018}\natexlab{}.
\newblock \showarticletitle{Artwork Personalization at Netflix}. In
  \bibinfo{booktitle}{\emph{Proceedings of the 12th ACM Conference on
  Recommender Systems}} \emph{(\bibinfo{series}{RecSys '18})}.
  \bibinfo{publisher}{ACM}, \bibinfo{address}{New York, NY, USA},
  \bibinfo{pages}{487--488}.
\newblock
\showISBNx{978-1-4503-5901-6}
\urldef\tempurl%
\url{https://doi.org/10.1145/3240323.3241729}
\showDOI{\tempurl}


\bibitem[\protect\citeauthoryear{Bradski}{Bradski}{2000}]%
        {opencv}
\bibfield{author}{\bibinfo{person}{G. Bradski}.}
  \bibinfo{year}{2000}\natexlab{}.
\newblock \showarticletitle{{The OpenCV Library}}.
\newblock \bibinfo{journal}{\emph{Dr. Dobb's Journal of Software Tools}}
  (\bibinfo{year}{2000}).
\newblock


\bibitem[\protect\citeauthoryear{Chatfield, Simonyan, Vedaldi, and
  Zisserman}{Chatfield et~al\mbox{.}}{2014}]%
        {vgg}
\bibfield{author}{\bibinfo{person}{Ken Chatfield}, \bibinfo{person}{Karen
  Simonyan}, \bibinfo{person}{Andrea Vedaldi}, {and} \bibinfo{person}{Andrew
  Zisserman}.} \bibinfo{year}{2014}\natexlab{}.
\newblock \showarticletitle{Return of the Devil in the Details: Delving Deep
  into Convolutional Nets}.
\newblock \bibinfo{journal}{\emph{CoRR}}  \bibinfo{volume}{abs/1405.3531}
  (\bibinfo{year}{2014}).
\newblock
\showeprint[arxiv]{1405.3531}
\urldef\tempurl%
\url{http://arxiv.org/abs/1405.3531}
\showURL{%
\tempurl}


\bibitem[\protect\citeauthoryear{Chen, Sun, Li, Lu, and Hua}{Chen
  et~al\mbox{.}}{2016}]%
        {deepctr}
\bibfield{author}{\bibinfo{person}{Junxuan Chen}, \bibinfo{person}{Baigui Sun},
  \bibinfo{person}{Hao Li}, \bibinfo{person}{Hongtao Lu}, {and}
  \bibinfo{person}{Xian-Sheng Hua}.} \bibinfo{year}{2016}\natexlab{}.
\newblock \showarticletitle{Deep ctr prediction in display advertising}. In
  \bibinfo{booktitle}{\emph{Proceedings of the 24th ACM international
  conference on Multimedia}}. ACM, \bibinfo{pages}{811--820}.
\newblock


\bibitem[\protect\citeauthoryear{Esfandarani and Milanfar}{Esfandarani and
  Milanfar}{2017}]%
        {nima_google}
\bibfield{author}{\bibinfo{person}{Hossein~Talebi Esfandarani} {and}
  \bibinfo{person}{Peyman Milanfar}.} \bibinfo{year}{2017}\natexlab{}.
\newblock \showarticletitle{{NIMA:} Neural Image Assessment}.
\newblock \bibinfo{journal}{\emph{CoRR}}  \bibinfo{volume}{abs/1709.05424}
  (\bibinfo{year}{2017}).
\newblock
\showeprint[arxiv]{1709.05424}
\urldef\tempurl%
\url{http://arxiv.org/abs/1709.05424}
\showURL{%
\tempurl}


\bibitem[\protect\citeauthoryear{{Fire} and {Schler}}{{Fire} and
  {Schler}}{2015}]%
        {explore_ads}
\bibfield{author}{\bibinfo{person}{Michael {Fire}} {and}
  \bibinfo{person}{Jonathan {Schler}}.} \bibinfo{year}{2015}\natexlab{}.
\newblock \showarticletitle{{Exploring Online Ad Images Using a Deep
  Convolutional Neural Network Approach}}.
\newblock \bibinfo{journal}{\emph{arXiv e-prints}}, Article
  \bibinfo{articleno}{arXiv:1509.00568} (\bibinfo{date}{Sep}
  \bibinfo{year}{2015}), \bibinfo{numpages}{arXiv:1509.00568}~pages.
\newblock
\showeprint[arxiv]{cs.CV/1509.00568}


\bibitem[\protect\citeauthoryear{Fortin, {De Rainville}, Gardner, Parizeau, and
  Gagn\'e}{Fortin et~al\mbox{.}}{2012}]%
        {DEAP_JMLR2012}
\bibfield{author}{\bibinfo{person}{F\'elix-Antoine Fortin},
  \bibinfo{person}{Fran\c{c}ois-Michel {De Rainville}},
  \bibinfo{person}{Marc-Andr\'e Gardner}, \bibinfo{person}{Marc Parizeau},
  {and} \bibinfo{person}{Christian Gagn\'e}.} \bibinfo{year}{2012}\natexlab{}.
\newblock \showarticletitle{{DEAP}: Evolutionary Algorithms Made Easy}.
\newblock \bibinfo{journal}{\emph{Journal of Machine Learning Research}}
  \bibinfo{volume}{13} (\bibinfo{date}{jul} \bibinfo{year}{2012}),
  \bibinfo{pages}{2171--2175}.
\newblock


\bibitem[\protect\citeauthoryear{Girshick, Donahue, Darrell, and
  Malik}{Girshick et~al\mbox{.}}{2014}]%
        {rcnn}
\bibfield{author}{\bibinfo{person}{Ross Girshick}, \bibinfo{person}{Jeff
  Donahue}, \bibinfo{person}{Trevor Darrell}, {and} \bibinfo{person}{Jitendra
  Malik}.} \bibinfo{year}{2014}\natexlab{}.
\newblock \showarticletitle{Rich feature hierarchies for accurate object
  detection and semantic segmentation}. In
  \bibinfo{booktitle}{\emph{Proceedings of the IEEE conference on computer
  vision and pattern recognition}}. \bibinfo{pages}{580--587}.
\newblock


\bibitem[\protect\citeauthoryear{{He}, {Gkioxari}, {Doll{\'a}r}, and
  {Girshick}}{{He} et~al\mbox{.}}{2017}]%
        {maskrcnn}
\bibfield{author}{\bibinfo{person}{Kaiming {He}}, \bibinfo{person}{Georgia
  {Gkioxari}}, \bibinfo{person}{Piotr {Doll{\'a}r}}, {and}
  \bibinfo{person}{Ross {Girshick}}.} \bibinfo{year}{2017}\natexlab{}.
\newblock \showarticletitle{{Mask R-CNN}}.
\newblock \bibinfo{journal}{\emph{arXiv e-prints}}, Article
  \bibinfo{articleno}{arXiv:1703.06870} (\bibinfo{date}{Mar}
  \bibinfo{year}{2017}), \bibinfo{numpages}{arXiv:1703.06870}~pages.
\newblock
\showeprint[arxiv]{cs.CV/1703.06870}


\bibitem[\protect\citeauthoryear{Hua}{Hua}{2018}]%
        {luban}
\bibfield{author}{\bibinfo{person}{Xian-Sheng Hua}.}
  \bibinfo{year}{2018}\natexlab{}.
\newblock \showarticletitle{Challenges and Practices of Large Scale Visual
  Intelligence in the Real-World}. In \bibinfo{booktitle}{\emph{Proceedings of
  the 26th ACM International Conference on Multimedia}}
  \emph{(\bibinfo{series}{MM '18})}. \bibinfo{publisher}{ACM},
  \bibinfo{address}{New York, NY, USA}, \bibinfo{pages}{364--364}.
\newblock
\showISBNx{978-1-4503-5665-7}
\urldef\tempurl%
\url{https://doi.org/10.1145/3240508.3267342}
\showDOI{\tempurl}


\bibitem[\protect\citeauthoryear{Jahanian, Liu, Lin, Tretter, O'Brien-Strain,
  Lee, Lyons, and Allebach}{Jahanian et~al\mbox{.}}{2013}]%
        {magazinedesign}
\bibfield{author}{\bibinfo{person}{Ali Jahanian}, \bibinfo{person}{Jerry Liu},
  \bibinfo{person}{Qian Lin}, \bibinfo{person}{Daniel Tretter},
  \bibinfo{person}{Eamonn O'Brien-Strain}, \bibinfo{person}{Seungyon~Claire
  Lee}, \bibinfo{person}{Nic Lyons}, {and} \bibinfo{person}{Jan Allebach}.}
  \bibinfo{year}{2013}\natexlab{}.
\newblock \showarticletitle{Recommendation system for automatic design of
  magazine covers}. In \bibinfo{booktitle}{\emph{Proceedings of the 2013
  international conference on Intelligent user interfaces}}. ACM,
  \bibinfo{pages}{95--106}.
\newblock


\bibitem[\protect\citeauthoryear{J{\"a}rvelin and
  Kek{\"a}l{\"a}inen}{J{\"a}rvelin and Kek{\"a}l{\"a}inen}{2002}]%
        {ndcg}
\bibfield{author}{\bibinfo{person}{Kalervo J{\"a}rvelin} {and}
  \bibinfo{person}{Jaana Kek{\"a}l{\"a}inen}.} \bibinfo{year}{2002}\natexlab{}.
\newblock \showarticletitle{Cumulated gain-based evaluation of IR techniques}.
\newblock \bibinfo{journal}{\emph{ACM Transactions on Information Systems
  (TOIS)}} \bibinfo{volume}{20}, \bibinfo{number}{4} (\bibinfo{year}{2002}),
  \bibinfo{pages}{422--446}.
\newblock


\bibitem[\protect\citeauthoryear{K.F.~Man and Kwong}{K.F.~Man and
  Kwong}{1996}]%
        {Genetic_algo_paper}
\bibfield{author}{\bibinfo{person}{K.S.~Tang K.F.~Man} {and}
  \bibinfo{person}{S. Kwong}.} \bibinfo{year}{1996}\natexlab{}.
\newblock \showarticletitle{Genetic Algolithms: Concepts and Applications}.
\newblock \bibinfo{journal}{\emph{IEEE Transactions on Industrial Electronics}}
  \bibinfo{volume}{43}, \bibinfo{number}{5} (\bibinfo{year}{1996}),
  \bibinfo{pages}{519 -- 534}.
\newblock


\bibitem[\protect\citeauthoryear{Li, Xu, Zhang, Hertzmann, and Yang}{Li
  et~al\mbox{.}}{2019}]%
        {layoutgan}
\bibfield{author}{\bibinfo{person}{Jianan Li}, \bibinfo{person}{Tingfa Xu},
  \bibinfo{person}{Jianming Zhang}, \bibinfo{person}{Aaron Hertzmann}, {and}
  \bibinfo{person}{Jimei Yang}.} \bibinfo{year}{2019}\natexlab{}.
\newblock \showarticletitle{Layout{GAN}: Generating Graphic Layouts with
  Wireframe Discriminator}. In \bibinfo{booktitle}{\emph{International
  Conference on Learning Representations}}.
\newblock
\urldef\tempurl%
\url{https://openreview.net/forum?id=HJxB5sRcFQ}
\showURL{%
\tempurl}


\bibitem[\protect\citeauthoryear{Maheshwari, Bansal, Dwivedi, Kumar, Manerikar,
  and Srinivasan}{Maheshwari et~al\mbox{.}}{2019}]%
        {adobepaper}
\bibfield{author}{\bibinfo{person}{Paridhi Maheshwari}, \bibinfo{person}{Nitish
  Bansal}, \bibinfo{person}{Surya Dwivedi}, \bibinfo{person}{Rohan Kumar},
  \bibinfo{person}{Pranav Manerikar}, {and} \bibinfo{person}{Balaji~Vasan
  Srinivasan}.} \bibinfo{year}{2019}\natexlab{}.
\newblock \showarticletitle{Exemplar Based Experience Transfer}. In
  \bibinfo{booktitle}{\emph{Proceedings of the 24th International Conference on
  Intelligent User Interfaces}} \emph{(\bibinfo{series}{IUI '19})}.
  \bibinfo{publisher}{ACM}, \bibinfo{address}{New York, NY, USA},
  \bibinfo{pages}{673--680}.
\newblock
\showISBNx{978-1-4503-6272-6}
\urldef\tempurl%
\url{https://doi.org/10.1145/3301275.3302300}
\showDOI{\tempurl}


\bibitem[\protect\citeauthoryear{O’Donovan, Agarwala, and
  Hertzmann}{O’Donovan et~al\mbox{.}}{2014}]%
        {single_page_layout}
\bibfield{author}{\bibinfo{person}{Peter O’Donovan}, \bibinfo{person}{Aseem
  Agarwala}, {and} \bibinfo{person}{Aaron Hertzmann}.}
  \bibinfo{year}{2014}\natexlab{}.
\newblock \showarticletitle{Learning layouts for single-pagegraphic designs}.
\newblock \bibinfo{journal}{\emph{IEEE transactions on visualization and
  computer graphics}} \bibinfo{volume}{20}, \bibinfo{number}{8}
  (\bibinfo{year}{2014}), \bibinfo{pages}{1200--1213}.
\newblock


\bibitem[\protect\citeauthoryear{Ren, He, Girshick, and Sun}{Ren
  et~al\mbox{.}}{2015}]%
        {fasterrcnn}
\bibfield{author}{\bibinfo{person}{Shaoqing Ren}, \bibinfo{person}{Kaiming He},
  \bibinfo{person}{Ross Girshick}, {and} \bibinfo{person}{Jian Sun}.}
  \bibinfo{year}{2015}\natexlab{}.
\newblock \showarticletitle{Faster r-cnn: Towards real-time object detection
  with region proposal networks}. In \bibinfo{booktitle}{\emph{Advances in
  neural information processing systems}}. \bibinfo{pages}{91--99}.
\newblock


\bibitem[\protect\citeauthoryear{Sandhaus, Rabbath, and Boll}{Sandhaus
  et~al\mbox{.}}{2011}]%
        {photobook}
\bibfield{author}{\bibinfo{person}{Philipp Sandhaus}, \bibinfo{person}{Mohammad
  Rabbath}, {and} \bibinfo{person}{Susanne Boll}.}
  \bibinfo{year}{2011}\natexlab{}.
\newblock \showarticletitle{Employing aesthetic principles for automatic photo
  book layout}. In \bibinfo{booktitle}{\emph{International Conference on
  Multimedia Modeling}}. Springer, \bibinfo{pages}{84--95}.
\newblock


\bibitem[\protect\citeauthoryear{Uchida}{Uchida}{2018}]%
        {gender_detection}
\bibfield{author}{\bibinfo{person}{Yusuke Uchida}.}
  \bibinfo{year}{2018}\natexlab{}.
\newblock \bibinfo{title}{Age-Gender-Estimation}.
\newblock
\newblock
\urldef\tempurl%
\url{https://github.com/yu4u/age-gender-estimation}
\showURL{%
\tempurl}


\bibitem[\protect\citeauthoryear{Xu}{Xu}{2017}]%
        {lubanBlog}
\bibfield{author}{\bibinfo{person}{Rexroth Xu}.}
  \bibinfo{year}{2017}\natexlab{}.
\newblock \bibinfo{title}{AI visual design is already here}.
\newblock
\newblock
\urldef\tempurl%
\url{https://medium.com/@rexrothX/}
\showURL{%
\tempurl}


\bibitem[\protect\citeauthoryear{Zhang, Hu, Ren, Yang, Xu, and Hua}{Zhang
  et~al\mbox{.}}{2017}]%
        {layout_alibaba_paper}
\bibfield{author}{\bibinfo{person}{Yunke Zhang}, \bibinfo{person}{Kangkang Hu},
  \bibinfo{person}{Peiran Ren}, \bibinfo{person}{Changyuan Yang},
  \bibinfo{person}{Weiwei Xu}, {and} \bibinfo{person}{Xian-Sheng Hua}.}
  \bibinfo{year}{2017}\natexlab{}.
\newblock \showarticletitle{Layout Style Modeling for Automating Banner
  Design}. In \bibinfo{booktitle}{\emph{Proceedings of the on Thematic
  Workshops of ACM Multimedia 2017}}. ACM, \bibinfo{pages}{451--459}.
\newblock


\bibitem[\protect\citeauthoryear{Zhou, Lapedriza, Xiao, Torralba, and
  Oliva}{Zhou et~al\mbox{.}}{2014}]%
        {scene_paper}
\bibfield{author}{\bibinfo{person}{Bolei Zhou}, \bibinfo{person}{Agata
  Lapedriza}, \bibinfo{person}{Jianxiong Xiao}, \bibinfo{person}{Antonio
  Torralba}, {and} \bibinfo{person}{Aude Oliva}.}
  \bibinfo{year}{2014}\natexlab{}.
\newblock \showarticletitle{Learning deep features for scene recognition using
  places database}. In \bibinfo{booktitle}{\emph{Advances in neural information
  processing systems}}. \bibinfo{pages}{487--495}.
\newblock


\bibitem[\protect\citeauthoryear{Zhou, Redi, Haines, and Lalmas}{Zhou
  et~al\mbox{.}}{2016}]%
        {nativequality}
\bibfield{author}{\bibinfo{person}{Ke Zhou}, \bibinfo{person}{Miriam Redi},
  \bibinfo{person}{Andrew Haines}, {and} \bibinfo{person}{Mounia Lalmas}.}
  \bibinfo{year}{2016}\natexlab{}.
\newblock \showarticletitle{Predicting pre-click quality for native
  advertisements}. In \bibinfo{booktitle}{\emph{Proceedings of the 25th
  International Conference on World Wide Web}}. International World Wide Web
  Conferences Steering Committee, \bibinfo{pages}{299--310}.
\newblock


\end{thebibliography}
\balance

%
\onecolumn
\appendix
\section{More qualitative results}
Few examples showing the raw photo-shoot, results using baseline approach and our approach. We can see how faces, body parts can be cut or overlapped with design elements if baseline approach was used. 


\begin{table*}[ht]

  \caption{Few Qualitative results with baseline and our approach} \label{tab:baseline}
  \begin{tabular}{lll} \hline 
      \textbf{Photoshoot Image} & \textbf{Baseline approach }&  \textbf{Our approach}\\
      \hline  
      \includegraphics[width=1.45in]{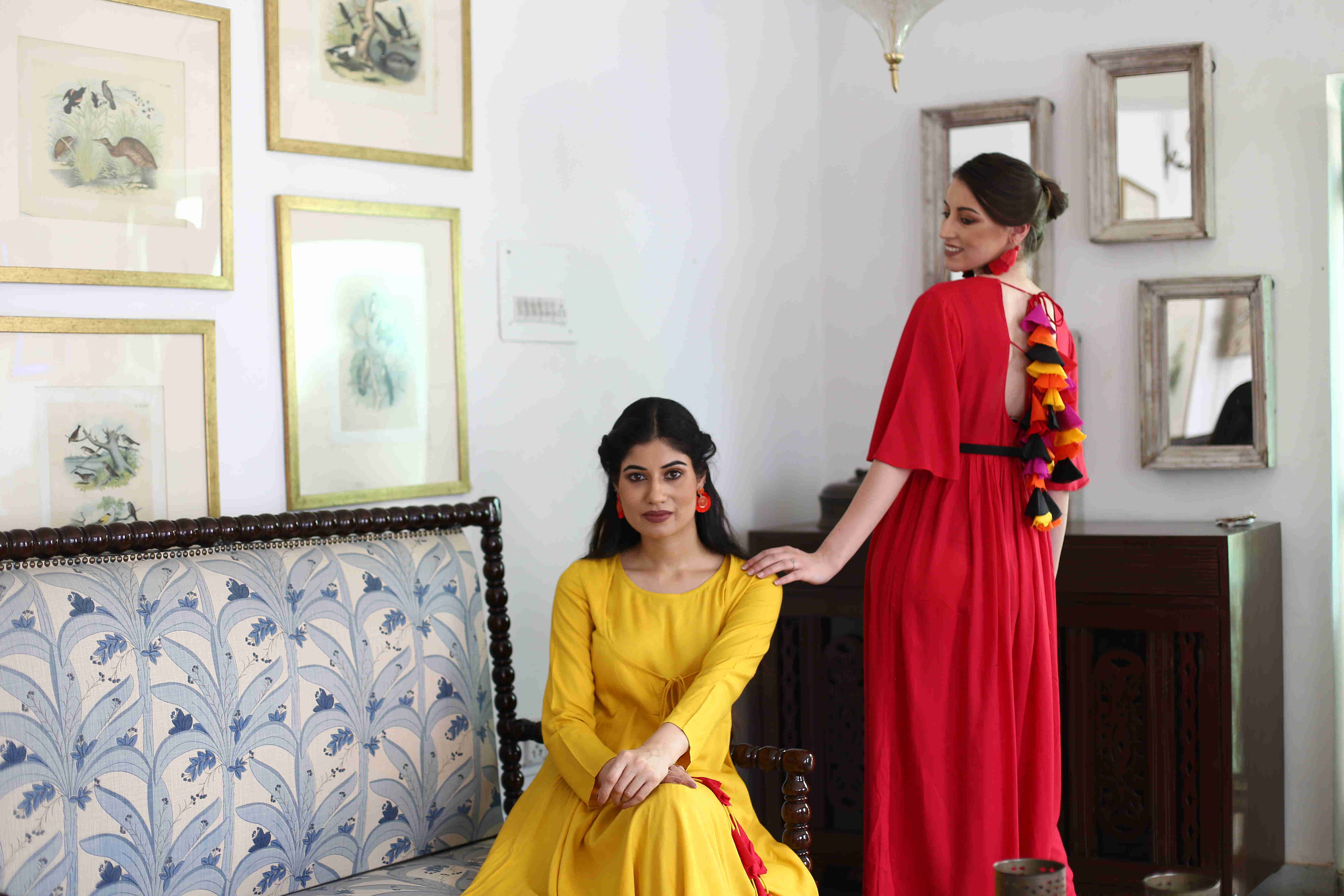} & \includegraphics[width=1.45in]{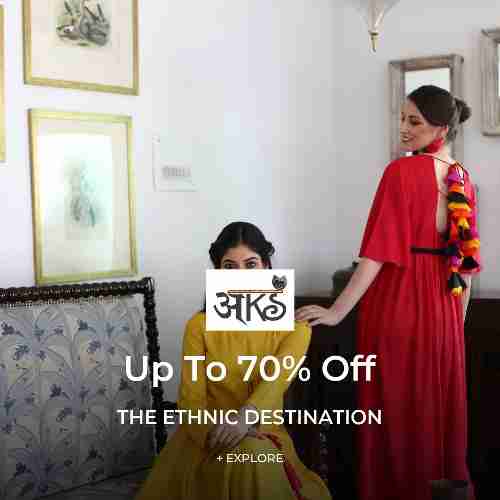} & \includegraphics[width=1.45in]{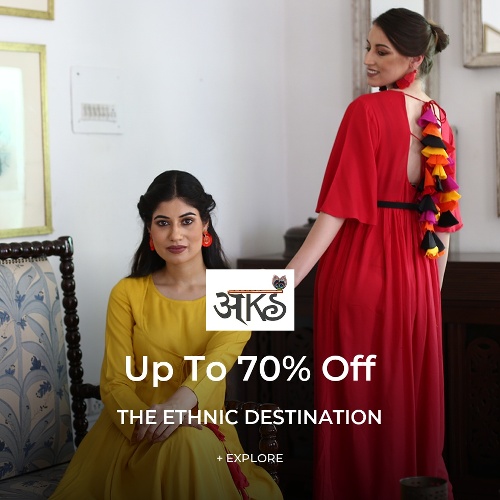} \\
      
      \includegraphics[width=1.45in]{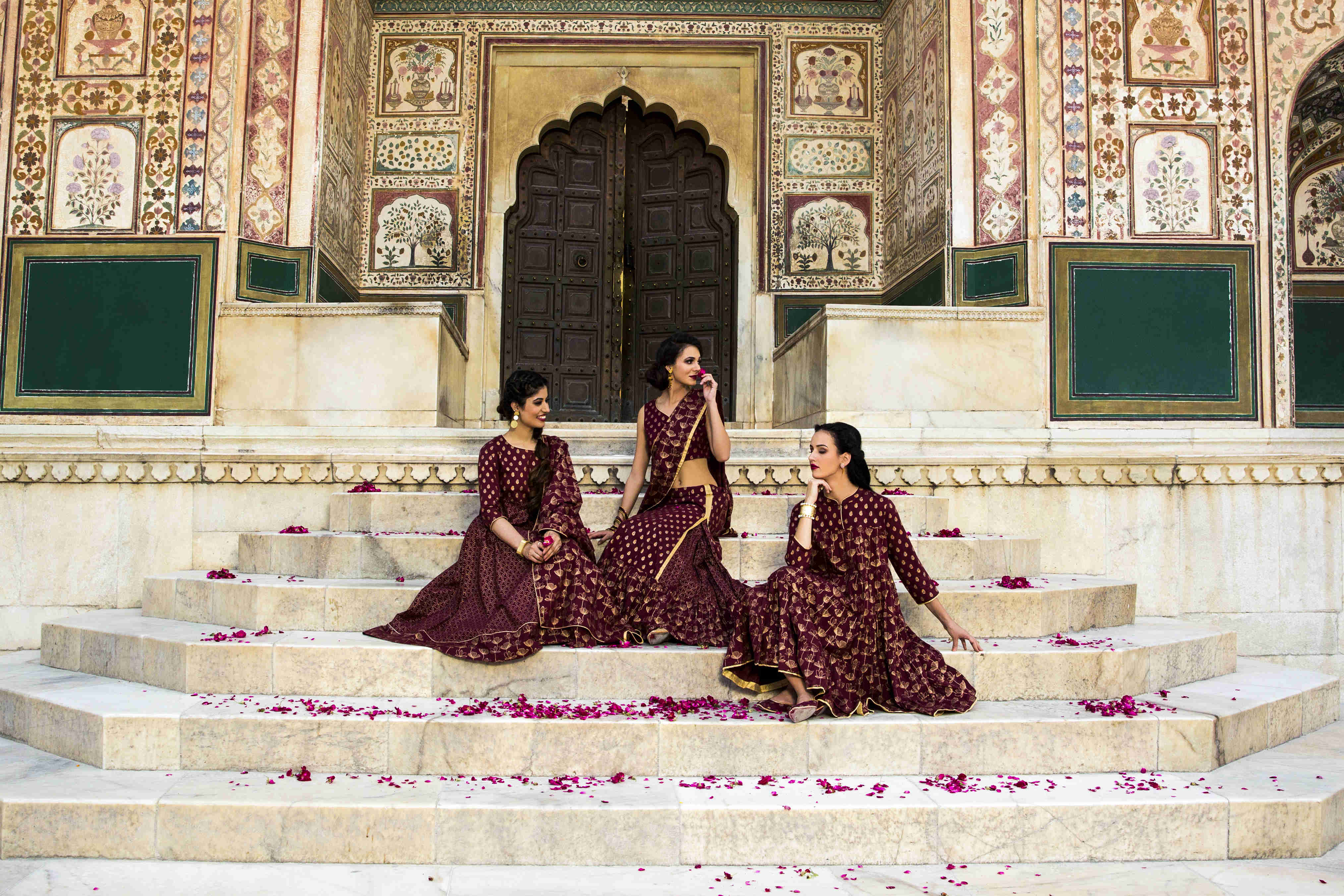} & \includegraphics[width=1.45in]{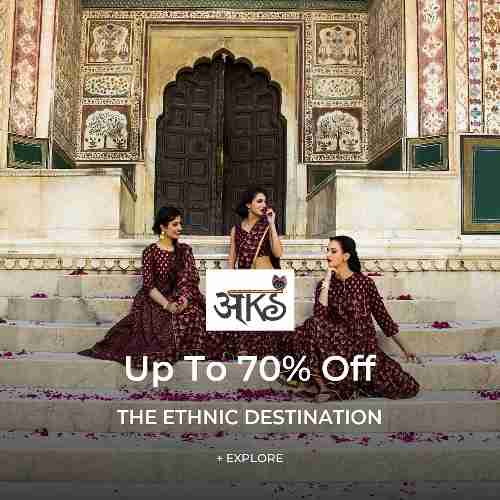} & \includegraphics[width=1.45in]{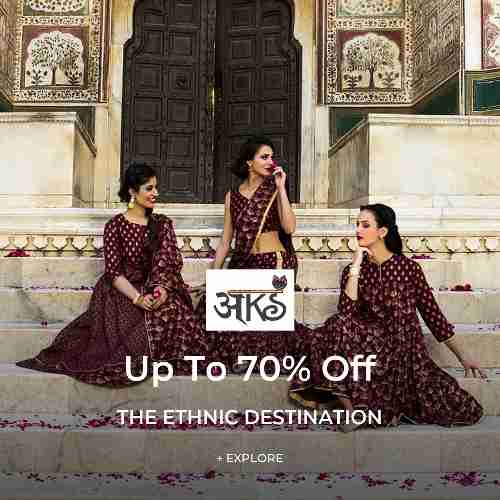} \\
      
      \includegraphics[width=1.45in]{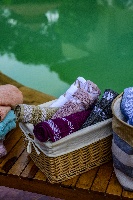} & \includegraphics[width=1.45in]{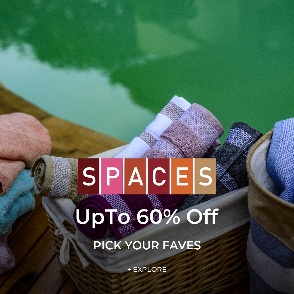} & \includegraphics[width=1.45in]{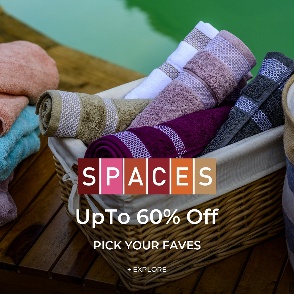} \\

      \includegraphics[width=1.45in]{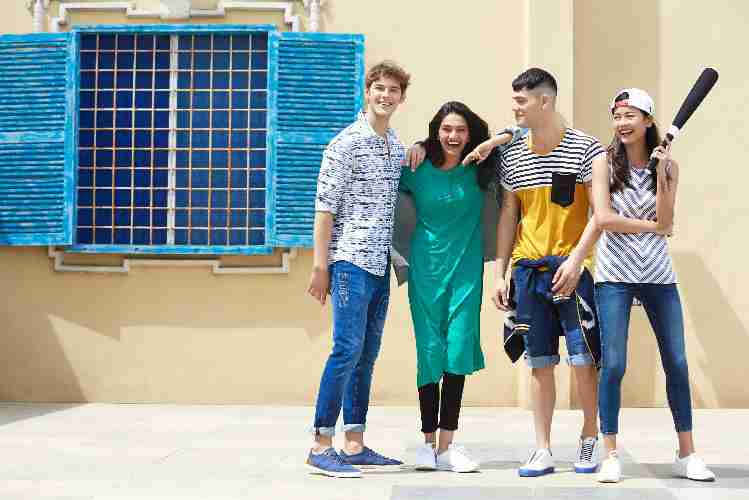} & \includegraphics[width=1.45in]{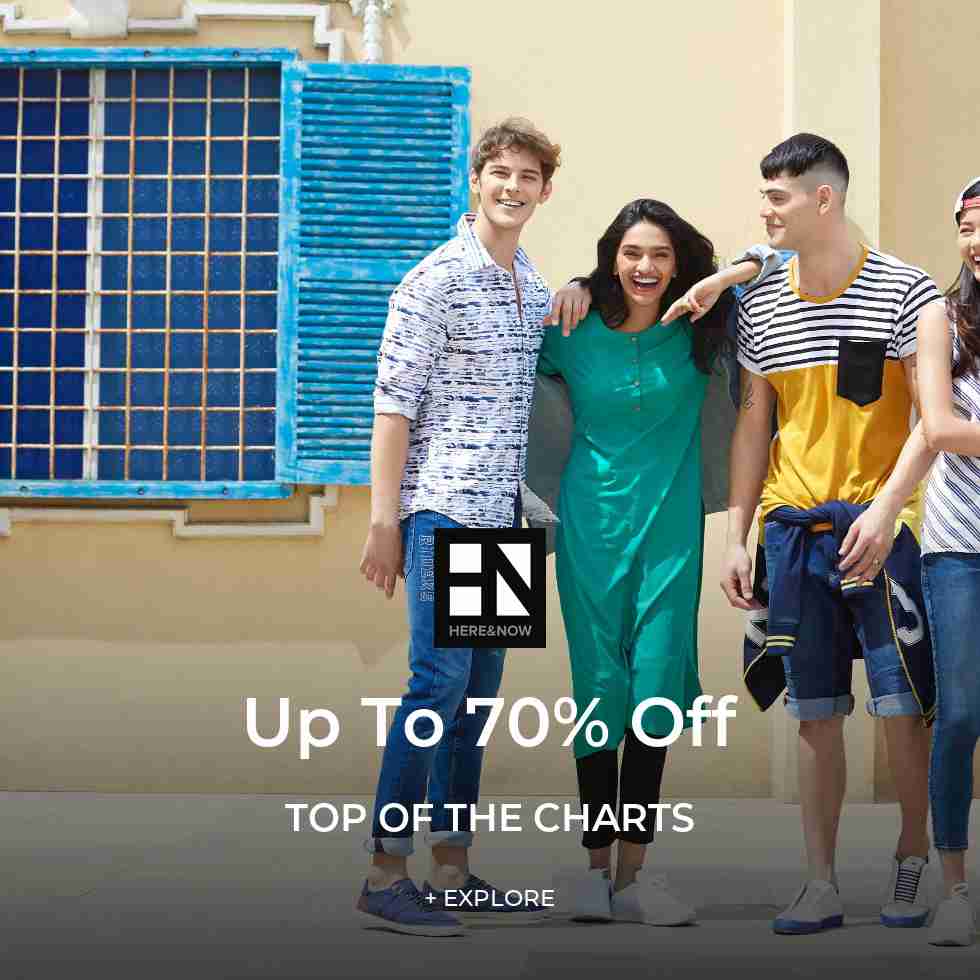} & \includegraphics[width=1.45in]{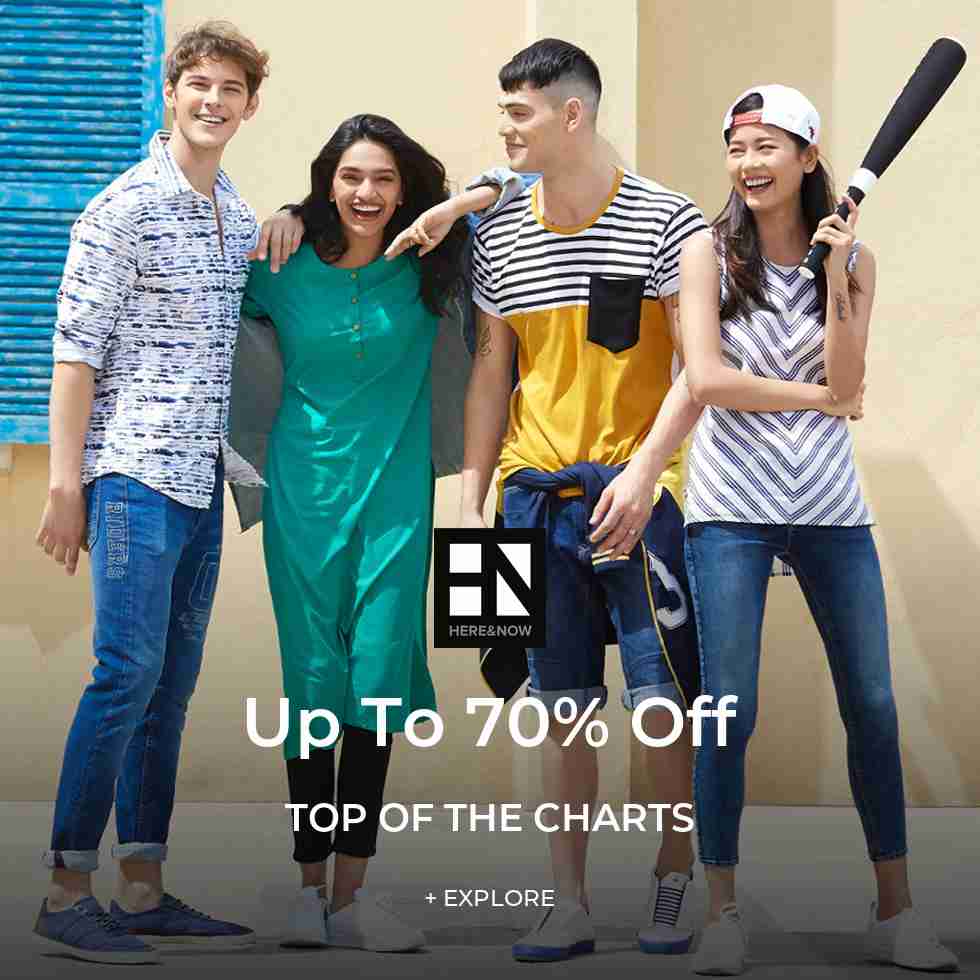} \\

      \hline
  \end{tabular}
\end{table*}


    
    
     
      

\end{document}